\documentclass[11pt,twoside]{article}
\usepackage{./asp2014}

\aspSuppressVolSlug
\resetcounters

\usepackage{color}

\newcommand{\eg}{e.\,g.}
\newcommand{\msun}{{\rm M_\odot}}

\begin{document}

\title{Supermassive Black Hole Pairs and Binaries}
\author{Sarah~Burke-Spolaor$^{1,2}$,
Laura Blecha$^3$,
Tamara Bogdanovi\'c$^{4}$,
Julia M. Comerford$^{5}$,
Joseph~Lazio$^6$,
Xin Liu$^7$,
Thomas~J.~Maccarone$^8$,
Dominic~Pesce$^{9,10}$,
Yue Shen$^7$,
and
Greg Taylor$^{11}$}
\affil{$^1$ Department of Physics and Astronomy, West Virginia University, Morgantown WV 26505, USA; \email{sarah.spolaor@mail.wvu.edu}}
\affil{$^2$ Center for Gravitational Waves and Cosmology, West Virginia University, Morgantown WV 26501, USA} 
\affil{$^3$ Department of Physics, University of Florida, Gainesville, FL 32611, USA;
\email{lblecha@ufl.edu}}
\affil{$^{4}$Center for Relativistic Astrophysics, Georgia Institute of Technology, Atlanta, GA 30332, USA;
\email{tamarab@gatech.edu}}
\affil{$^{5}$Department of Astrophysical and Planetary Sciences, University of Colorado, Boulder, CO 80309, USA;
\email{Julie.Comerford@colorado.edu}}
\affil{$^6$Jet Propulsion Laboratory, California Institute of Technology, Pasadena CA, 91109, USA}
\affil{$^7$Department of Astronomy, University of Illinois at Urbana-Champaign, Urbana, IL 61801, USA; \email{xinliuxl@illinois.edu, shenyue@illinois.edu}}
\affil{$^8$Department of Physics \& Astronomy, Texas Tech University, Lubbock, TX 79409-1051, USA}
\affil{$^9$Department of Astronomy, University of Virginia, 530 McCormick Road, Charlottesville, VA 22904, USA}
\affil{$^{10}$National Radio Astronomy Observatory, 520 Edgemont Road, Charlottesville, VA 22903, USA; \email{dpesce@virginia.edu}}
\affil{$^{11}$Department of Physics and Astronomy, University of New Mexico, Albuquerque, NM 87131, USA;
\email{gbtaylor@unm.edu}}

\paperauthor{Sample~Author1}{Author1Email@email.edu}{ORCID_Or_Blank}{Author1 Institution}{Author1 Department}{City}{State/Province}{Postal Code}{Country}
\paperauthor{Sample~Author2}{Author2Email@email.edu}{ORCID_Or_Blank}{Author2 Institution}{Author2 Department}{City}{State/Province}{Postal Code}{Country}
\paperauthor{Sample~Author3}{Author3Email@email.edu}{ORCID_Or_Blank}{Author3 Institution}{Author3 Department}{City}{State/Province}{Postal Code}{Country}

\section{The Importance and Impact of Binary Supermassive Black Hole Observations}\label{sec:intro}

The ngVLA will be a powerful probe of dual and binary supermassive black holes (SMBHs), performing detailed studies of their population and evolution, and enabling powerful multi-messenger science when combined with gravitational-wave detection. The ngVLA will enable such studies by an unprecedented combination of sensitivity, frequency coverage, and particularly if equipped with long baselines, essentially unparalleled angular resolution.

\begin{figure}
\centering
\includegraphics[width=0.90\textwidth,trim=0mm 0mm 0mm 0mm, clip]{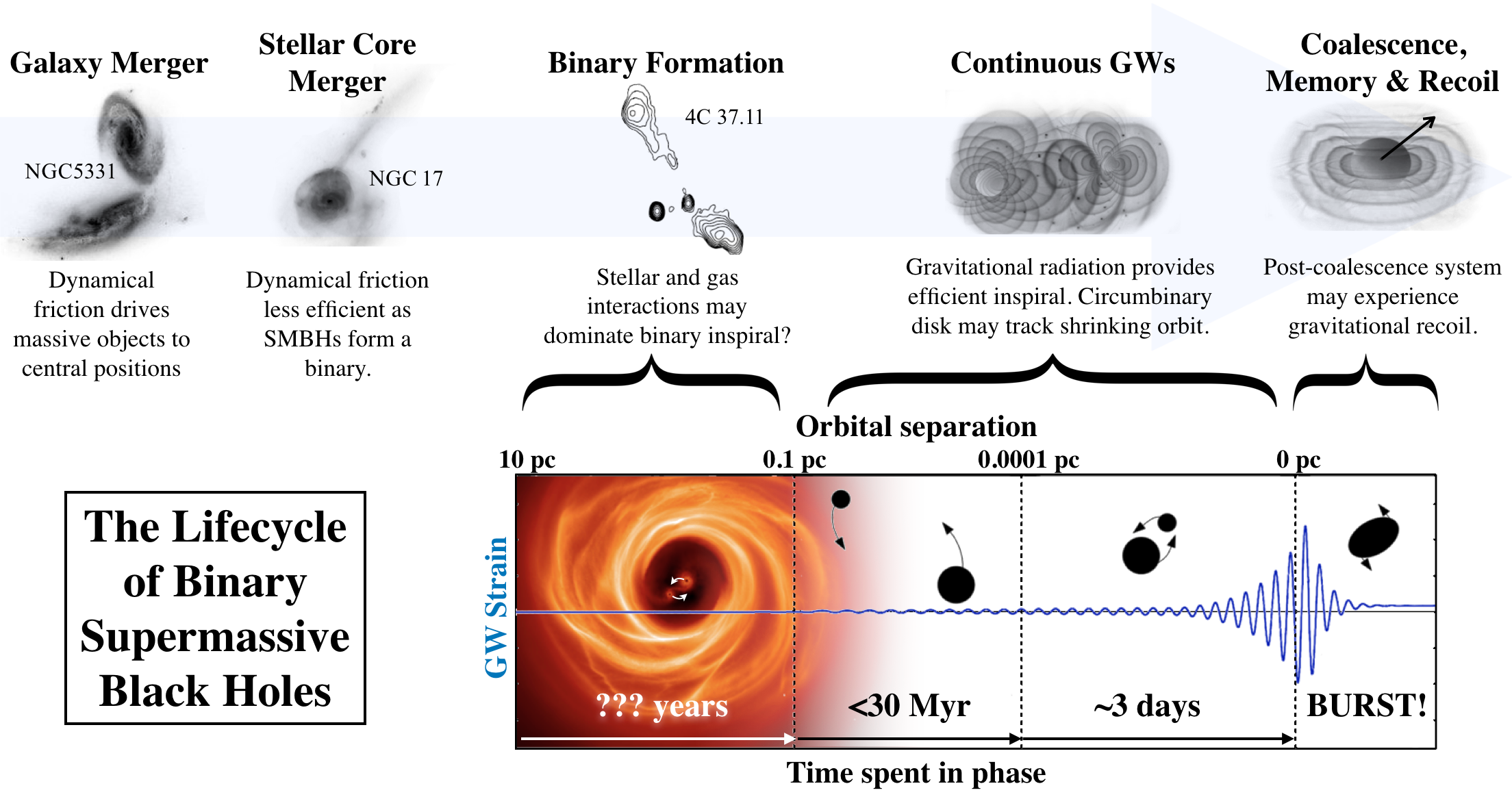}
\vspace{-4mm}
\caption{Figure from \citet{GWAWG-paper}, summarizing the binary SMBH life-cycle. A major unknown in binary evolution theory is the efficiency of inspiral from $\sim$10\,pc down to $\sim$0.1\,pc separations, after which the binary can coalesce efficiently due to gravitational waves.
The ngVLA will discover widely separated binaries, and depending on its long-baseline sensitivity, could discover
targets detectable by pulsar timing arrays (PTAs).
PTAs such as NANOGrav can detect supermassive ($>10^8\,\msun$) binaries within $\sim$0.1\,pc separation (second panel in the lower figure).
On rare occasion, PTAs may detect the permanent space-time deformation (GW memory) caused by a binary's coalescence \citep{favata09}. 
LISA will detect lower-mass binaries, up to $\sim10^7\,\msun$, in the weeks or months
leading up to coalescence.
Image credits: Galaxies, Hubble/STSci; 4C37.11, \citet{rodriguez}; Simulation visuals, C. Henze/NASA; Circumbinary accretion disk, C. Cuadra.
}\label{fig:lifecycle}
\vspace{-5mm}
\end{figure}

Dual (\mbox{$\lesssim$ 10\,kpc separation)} and binary (\mbox{$\lesssim$ 10\,pc} separation) SMBH systems can form during major galaxy mergers. 
Figure \ref{fig:lifecycle} illustrates the formation and evolution of double SMBH systems, from the initial stages of a merger of galaxies to the emission of gravitational waves to an SMBH merger \citep{begelman80}.
When one or both SMBHs power an active galactic nucleus, multi-wavelength emission can directly mark the presence and locations of the SMBHs. Jets produced by these active nuclei will generate radio emission. Because jet cores closely trace the location of a SMBH, and because of the extended lifetime of synchrotron jet observability, radio observations both at high resolution and of larger-scale jet structures provide excellent tools to trace the current and past dynamics of SMBHs in a merging system.

SMBH pairs confidently mark ongoing galaxy mergers and imminent SMBH coalescences, and are therefore a strong probe of redshift-dependent merger rates, occupation fractions of dual SMBHs in galaxies, and post-merger dynamical evolution. If we determine the rate and environments of systems containing two or more widely separated active galactic nuclei (AGN), we can explore merger-induced activity of nuclei and susbequent SMBH growth. These dual AGN probe the critical regime during which the black holes and their associated galactic-scale stellar cores are virialized (\eg\ within $\sim$10\,kpc black hole separation). During this phase, both star formation and AGN activity peak during galaxy mergers, according to cosmological simulations \citep{blecha13}. Having a sample of dual AGN is key to testing predictions from simulations, such as whether the most luminous AGN are preferentially triggered in mergers and whether there is a time-lag between star formation triggering and AGN triggering in mergers \citep{hopkins12,hopkins14}.

A systematic survey of dual and binary SMBHs also allows a direct prediction of the projected gravitational wave signals in the low and very low frequency regime of gravitational radiation. Both pulsar timing arrays (PTAs) and future space-based laser interferometers have binary SMBHs as a key target. They may detect both ``discrete'' individual targets, and a stochastic background of small-orbit SMBH binaries.

The ngVLA will also be coming online at a time when PTAs like the North American Nanohertz Observatory for Gravitational Waves (NANOGrav) expects to have already detected a number of discrete binary SMBHs \citep[\eg][]{mingarelli-2mass,kelley+submitted}. LISA is expected to directly constrain SMBH binary coalescence rates soon after its launch \citep{lisa-science-2,as+17}. 
Gravitational-wave and multi-messenger astronomy with SMBHs will thus be a leading pursuit of the ngVLA era. Particularly with long (VLBI) baselines, ngVLA will have the potential to identify the hosts of emitting gravitational-wave systems, enabling broad-scoped science that includes:
\begin{itemize}
\vspace{-2mm}\item Precise orbital tracking of binary SMBHs via parsec-scale jet morphology and core tracking \citep{bansal}.
\vspace{-2mm}\item Use of SMBHs as a standard ruler \citep{dorazioloeb}.
\vspace{-2mm}\item Measurement of merger-induced accretion rates and imaging of small-scale AGN feedback \citep{mullersanchez15}
\vspace{-2mm}\item Detailed multi-wavelength and multi-messenger studies of circumbinary disk evolution and binary inspiral timescales \citep[\eg][]{kelley+17}.
\vspace{-2mm}\item Calibration of SMBH-host relations up to moderate/high redshifts via direct SMBHB mass measurements (host identification breaks the distance/mass degeneracy encountered in PTA discrete source detection). 
\end{itemize}

In the remainder of this chapter, we describe the various approaches that can be used to identify SMBH pairs with the ngVLA (Sec.\,\ref{sec:search}), compare those with past searches in Sec.\,\ref{sec:past}, and discuss synergies with multi-wavelength observations in Sec.\,\ref{sec:synergies}. Finally, Sec.\,\ref{sec:ngvla} lays out several example ngVLA studies and their general requirements for success.

\section{Identifying Paired SMBHs with Radio Emission}\label{sec:search}
The ngVLA could be used to conduct surveys for SMBH pairs, follow up candidates identified in other surveys, or both. Either way, in order to reveal dual or binary SMBH candidates with the ngVLA, a few identification methods are available. ngVLA can identify pairs through studies of morphology, spectra, and/or time-dependence of the bulk relativistic flow of particles from the SMBHs.

\begin{figure}
\centering
\includegraphics[width=0.99\textwidth,trim=0mm 0mm 0mm 0mm, clip]{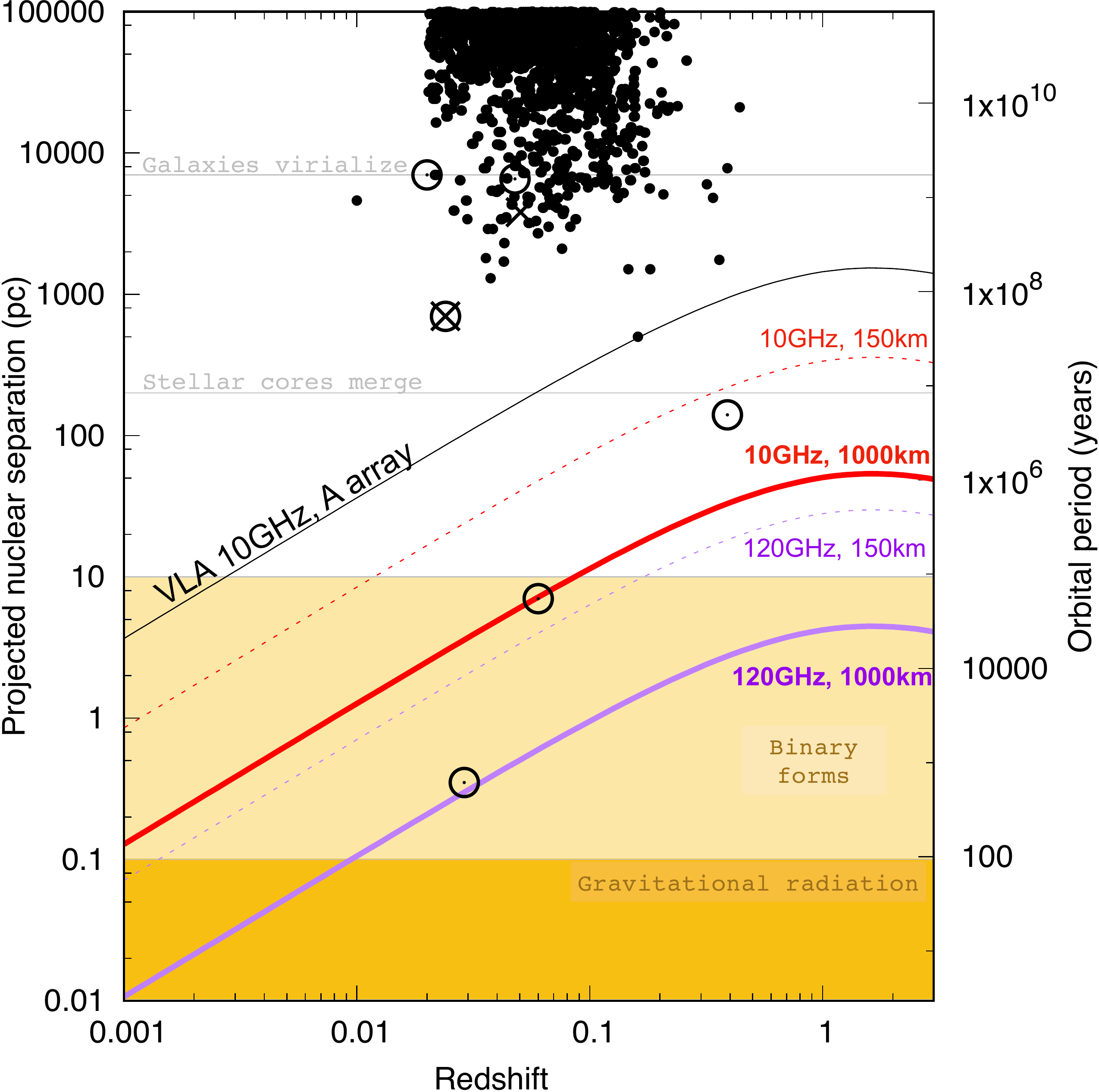}
\vspace{-2mm}
\caption{The observed population of candidate AGN pairs to date compared with the resolution of the ngVLA.  Each point represents a dual \hbox{AGN}
discovered via radio (X), X-ray (circles), or optical/near-IR (dots; from \citealt{liu+11}).
Approximate lines for critical stages in binary formation and evolution are marked as horizontal lines. 
The red, green, and purple curves indicate the resolution limit of 10\,GHz, 50\,GHz, and 120\,GHz center observing bands, respectively; for each observing set-up, one cannot resolve a binary orbit below that line. The dashed curves show the resolution limit of 150\,km baselines, while the solid curves show a nominal 1000\,km extended-baseline array. The ngVLA with a VLBI expansion can resolve, and hence directly image and identify, double supermassive black holes at sub-10\,pc separations. Longer baselines and higher frequencies have the potential to resolve multi-messenger SMBH binaries in the nearby Universe (indicated by the curves that cross the dark yellow region).}
\label{fig:vlares}
\end{figure}

\subsection{Direct Core Imaging}\label{sec:spectra}
Radio imaging of multiple self-absorbed synchrotron cores in the $\sim$1--50\,GHz frequency range is one of the most direct routes to identify dual SMBHs. It is exceedingly rare to find bright multiple flat-spectrum objects within close proximity \citep[\eg][]{rodriguez,radiocensus}. The radio ``core'' that is seen in SMBHs represents the shock front near the base of the radio jet, which is compact and in a region of considerable optical depth. These can show a flat spectrum (with spectral index $\alpha\gtrsim-0.5$, where $S\propto f^\alpha$) up to moderate GHz frequencies \citep[\eg][]{blandford+79}. This technique relies on detecting more than one flat-spectrum, compact radio source within a 
common merger host. Confident confirmation of the nature of such detections requires modelling of broad-band continuum spectra, high-resolution morphological modelling, and ideally long-term studies of proper motion. For instance, in the object 4C37.11 (Fig.\,\ref{fig:lifecycle}), there are two cores at a projected separation of 7\,pc that remain compact down to sub-mas resolutions, with larger-scale diffuse emission that support the interpretation of the two knots as distinct, but gravitationally interacting, SMBHs. Long-term astrometric observations of this object have led to the first tentative ``orbital tracking'' for a resolved binary, although its orbital period is fairly long at $\sim$$10^4$\,yr \citep{bansal}.

Figure~\ref{fig:vlares} shows a census of known and candidate resolved (imaged) dual SMBHs. Uniquely, radio identifications have probed separations within a few hundred parsecs: the radius of a typical galactic stellar bulge \citep{merrittLRR}. Double black hole systems with separations smaller than a few tens of parsecs are \emph{direct precursors} to the gravitational-wave emitters detectable by PTAs and space-based laser interferometers, so are of utmost interest to the target science outlined in Section\,\ref{sec:intro}. There is only one confident binary in this range \citep{rodriguez}, and only a scant list of candidates \citep{deane,kharb+17}.

Binary SMBHs are relatively rare, and performing a complete blind search for dual radio cores requires relatively high dynamic range and thorough $u, v$ coverage at high resolution. Present facilities have aspects of these but not all. While the VLA has excellent sensitivity and $u, v$ coverage, it lacks sufficient resolution to discover close pairs.
While current long-baseline facilities have excellent resolution, they lack the snapshot sensitivity and $u, v$ coverage to thoroughly probe cores \citep[\eg][]{deller+middelberg}. The ngVLA has a unique potential to reach a balance between these factors (Bansal et al.\ in prep).

\subsection{Jet Morphology}\label{sec:shapes}
\begin{figure}
\centering
\includegraphics[width=0.33\textwidth,trim=0mm 7mm 0mm 5mm,clip]{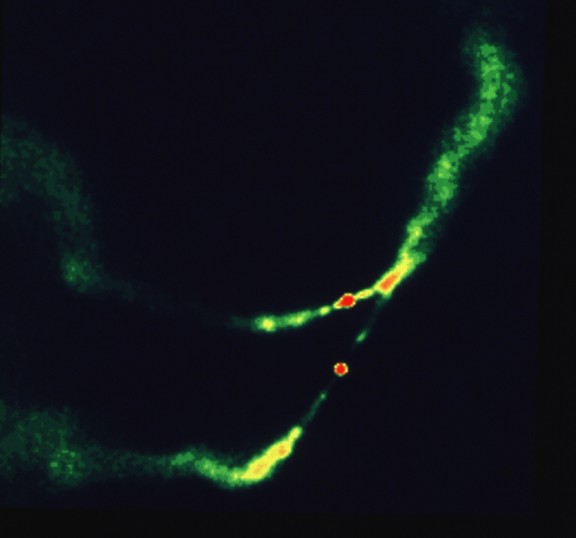}~
\includegraphics[width=0.33\textwidth,trim=0mm 15mm 0mm 10mm,clip]{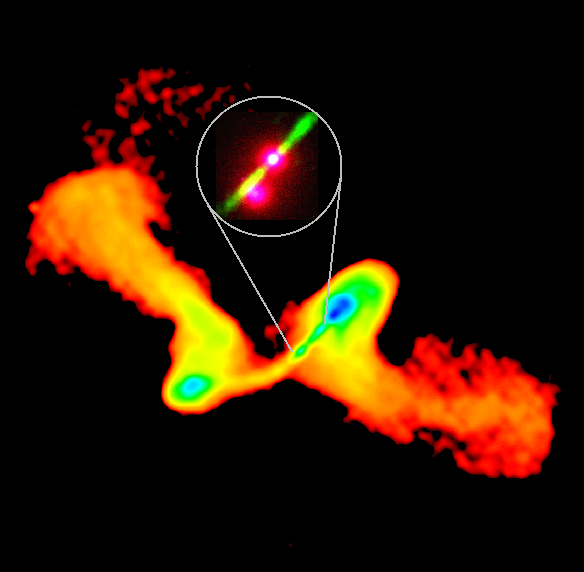}~
\includegraphics[width=0.2\textwidth]{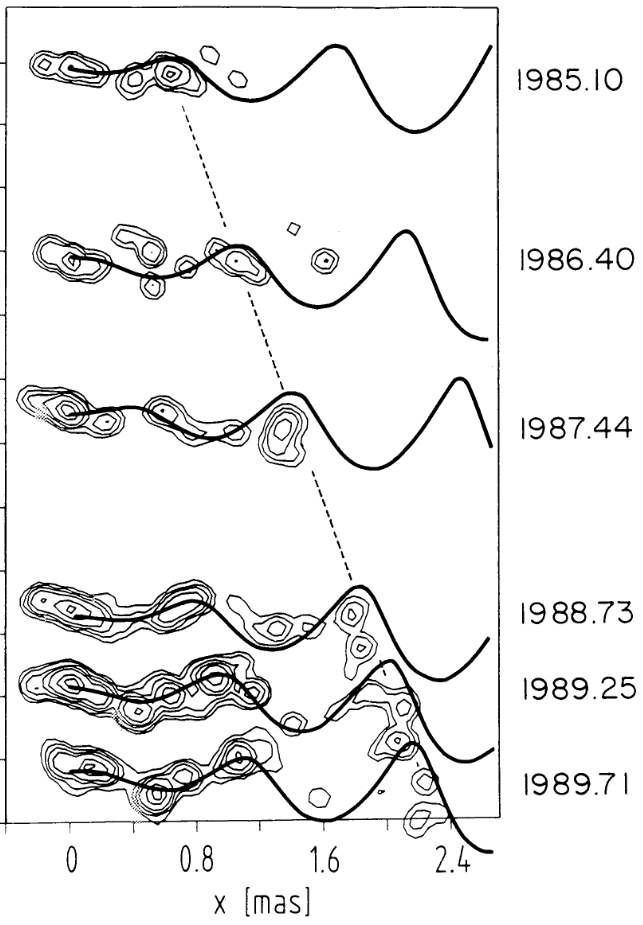}
\vspace{-2ex}
\caption{Three examples of candidate dual AGN identified through their static or time-dependent radio jet morphology (Sections \ref{sec:shapes}, \ref{sec:time}). \textit{Left:} 3C~75, the dual cores of which have a 7~kpc projected separation and exhibits large-scale jet structures that have misaligned but correlated morphology. (Credit: \hbox{NRAO/AUI}; F.~N.~Owen, C.~P.~O'Dea, M.~Inoue, \& J.~Eilek). \textit{Center:} NGC~326, which has an \texttt{X}-shaped structure that inspired its dual nucleus interpretation. The cores were confirmed at other wavelengths, as shown. (Credit: \hbox{NRAO/AUI}; STScI [inset]). \textit{Right:} The helical jet of 1928+738 shows a 2.9-year cycle \citep{roos+93}.}
\label{fig:dualagn}
\vspace{-3mm}
\end{figure}

Identification of dual AGN via large-scale jet morphology is somewhat fraught because of the difficulty in discerning external vs.\ internal influences on the jet(s).
Dual AGN emitting on long time scales can produce quadruple jets (Fig.~\ref{fig:dualagn}), and
binary interactions can result in precessing jets that produce large-scale S-shaped radio jet morphologies. Parabolic SMBH-SMBH encounters and binary coalescences can rapidly reorient a jet, potentially producing X-shaped morphologies \citep{merrittekers}. Previous large-scale radio surveys (\eg, \hbox{NVSS}, FIRST) have had a strong output of candidate binary or post-merger SMBHs identified through their large-scale jet morphologies. Recent surveys of such radio jet morphologies include 87 X-shaped radio jets found with the VLA \citep{roberts18}. However, past studies have revealed that many such morphologies can be described by backwards-flows of expelled jet material caused by fluid dynamics, rather than originating from a dual system \citep{xnotbinary}.

Still, some (as shown in Fig.\,\ref{fig:dualagn}) have been demonstrated either conclusively or extensively as candidate genuine dual AGN. Broad searches for such features require low-surface-brightness sensitivity coupled with resolution.

Large-scale periodic knots along radio jets have also been hypothesized to be periodic accretion episodes fueled by a binary; however, these are difficult to distinguish from flow instabilities \citep{godfrey_0637-752}.

\subsection{Variability in Morphology or Flux}\label{sec:time}
Some radio quasars show light curves that appear to be periodic or sinusoidal. At high spatial resolutions, some jets have morphological variability: e.\,g.\ 1928+738 has jets that trace out helical patterns as a function of time \citep[Figure\,\ref{fig:dualagn};][]{roos+93}.
A common interpretation of both of these effects is that an accreting, jet-producing black hole has a jet axis whose angle is being modified by a secondary, orbiting SMBH. Confident identification of a periodicity in a light curve requires long-term variability monitoring (over more than a few cycles) to ensure that the observed variability is genuinely cyclical, and not simply the detection of a red noise process \citep{vaughan+16,charisi+18}. Thus, light curve variability programs require a long-term time investment for candidate binary identification.

As with the tracking of 4C37.11 \citep{bansal}, in an ideal case we will be able to track the orbital movements of one or two distinct cores. However, making the simple but suitable assumption that binaries are circular and obey Kepler's third law, their approximate largest physical separation scales with binary period as \mbox{$a\propto P^{2/3}$}. This means that the majority of resolvable binaries will have orbital periods well beyond human lifetimes. Tracking the orbital motion will only be possible for long baselines, and for the massive compact systems at low redshifts.   
With long-baseline ($\sim$10,000\,km) arrays, such observations will be possible for $10^9 M\odot$ radio-emitting binaries with few parsec separations out to $z\sim0.1$ on timescales of a few years. This clearly
underscores the importance of a VLBI option for the ngVLA. If ngVLA baselines are limited to 300 km, then the timescales for resolving orbital properties will increase by a factor of 200, essentially making the ngVLA unsuitable for such studies despite its terrific snapshot sensitivity.


\subsection{Maser Dynamics}\label{sec:masers}
Long-baseline observations of masers have been instrumental in finding SMBH masses and analyzing accretion disk dynamics (e.g., \citealt{Miyoshi95}, \citealt{Braatz10}, \citealt{Kuo13}, \citealt{Gao16}).  Deviations from Keplerian motion in the maser rotation curves can indicate that the potential is not that of a point source (e.g., \citealt{Kuo11}), and they can therefore be sensitive to close SMBH binaries (i.e., orbital separations of roughly $\sim$0.01--0.1\,pc).

For wider SMBH binaries ($\sim$1--100\,pc separation), the maser rotation curve is not expected to be substantially different from Keplerian.  In this case, one can compare the barycenter velocity derived from the maser rotation curve (which unambiguously traces the SMBH velocity) to that of the galaxy on large scales ($\sim$10's of kpc, traced using stellar or HI dynamics). Significant differences between these two velocities are then indicative of binary or recoiling SMBHs, although a measurement of a velocity offset cannot by itself distinguish between the two possibilities (Pesce et al. 2018, submitted to ApJ). 




\section{Current Radio Searches and Limitations}\label{sec:past}
In the past, most radio-identified candidate binary SMBHs have mainly been found serendipitously.
A few concerted efforts have attempted to survey directly for binary SMBHs. Here we discuss these, their successes, and limitations.

A number of blind or semi-blind searches have been made for dual radio cores, which as described in Sec.\,\ref{sec:spectra} is most easily realized through spectral imaging to search for multiple compact, flat-spectrum radio components. \citet{radiocensus} searched all available historical data from the Very Long Baseline Array with dual-frequency imaging; despite the large sample size ($\sim$3200 AGN), this search was limited primarily by sensitivity.
The VIPS survey of radio cores \citep{VIPS} originally discovered 4C37.11 as an extended flat-spectrum object; this source was later evidenced to be a binary SMBH \citep{rodriguez}. \cite{tremblay16} performed multi-frequency imaging of the flux-limited $\sim$1100 AGN included in VIPS, all of which were pre-selected to be bright and flat-spectrum. This search did not identify any new binary systems from the VIPS sample, and concluded that 15 candidates remained unconfirmed. Finally, \citet{condon-2mass} performed deep observations of 834 radio-bright 2MASS galaxies, under the hypothesis that the most massive galaxies must have undergone a major merger in their history, and therefore might contain a binary or recoiling SMBH. So far from these surveys, only one resolved dual core, 4C37.11, was re-detected. On simple arguments of merger rates, inspiral timescales, and radio AGN luminosity functions, at least tens of SMBH pairs must exist within these samples; thus, it is clear that these past experiments have been limited by a combination of $u, v$ coverage, sensitivity, dynamic range, and perhaps sample size.

\section{Current Multi-wavelength Synergies and Limitations}\label{sec:synergies}
Dual and binary SMBH signatures have been proposed in a number of wavebands (see, \eg, the review in \citealt{sbs-cqg}), and searches pursuing those signatures have revealed a number of binary candidates. We note that in all cases SMBH binaries have been challenging to identify because of their small separation on the sky, the uncertainties related to the uniqueness of their observational signatures, and the limited baseline of monitoring campaigns. Direct radio imaging, however, can provide an excellent follow-up to explore the nature of objects identified with these signatures.

The most prolific binary search method thus far has been the use of photometric and spectroscopic optical signatures to reveal candidate SMBH binaries. We describe that work here given its relevance to upcoming synoptic sky surveys that will be contemporary to ngVLA.

Multi-wavelength (radio, optical, and X-ray) searches for SMBH systems with large separations, corresponding to early stages of galactic mergers, have so far successfully identified a number of dual and offset AGN, as shown in Fig.\,\ref{fig:dualagn} \citep[][and others]{ngc6240, fu11,koss11, koss16, liu13b, comerford15, barrows16,mullersanchez15}.

SMBHs with even smaller (parsec and sub-parsec) separations are representative of the later stages of galactic mergers in which the two SMBHs are sufficiently close to form a gravitationally bound pair. Direct imaging using VLBI can confidently detect such a system, however requires deep observations to do so successfully. In this regard, optical surveys can help to identify a sample of SMBH binaries for VLBI followup \citep[see][for a review]{Bogdanovic2015}.
Current studies have been exploring periodic quasar variability as a potential indicator for binary SMBH emissions; these and future surveys of this kind using synoptic instruments like PAN-STARRS and LSST will reveal many candidate binary systems ideal for follow-up with the ngVLA plus a long-baseline extension.
%

Current photometric surveys
have already uncovered about 150 SMBH binary candidates \citep[e.g.,][]{valtonen08, Graham2015, charisi16, liu16}. In this approach, the quasi-periodic variability in the lightcurves of monitored quasars is interpreted as a manifestation of binary orbital motion. Because of the finite temporal extent of the surveys, 
most of these candidates have orbital periods of order a few years, hence would be in the regime of gravitational-radiation-dominated inspiral (Fig.\,\ref{fig:lifecycle}). However, studies of this type suffer the same red-noise issues noted in Sec.\,\ref{sec:time}.
A recent, initially quite promising binary SMBH candidate from a quasar light curve, PG1302, has recently been losing traction. Additional data from the All-Sky Automated Survey for Supernovae (ASAS-SN), extending the light curve to present-day, shows that the periodicity does not persist and disfavors the binary SMBH interpretation \citep{liu18}.
Furthermore, the upper limit placed by PTAs on the gravitational wave background largely rules out the amplitude implied by the $\sim150$ photometric binary candidates, implying that some significant fraction of them are unlikely to be SMBH binaries \citep{sesana17}. While this will lead to a downward revision in the number of photometrically identified SMBH binary candidates, it provides a nice example of the effectiveness of multi-messenger techniques, when they can be combined.

Spectroscopic searches for SMBH binaries have so far identified several dozen candidates. They rely on the detection of a velocity shift in the emission-line spectrum of an SMBH binary that arises as a consequence of the binary orbital motion
\citep{gaskell83,gaskell96,bogdanovic09}. The main complication of this approach is that the velocity-shift signature is not unique to SMBH binaries; it may also arise from outflowing winds near the SMBH, or even from a gravitational-wave recoil kick \citep[e.g.,][]{popovic12,Eracleous2012,barth15}. To address this ambiguity, spectroscopic searches have been designed to monitor the offset of broad emission-line profiles over multiple epochs and target sources in which modulations in the offset are consistent with binary orbital motion \citep{bon12, bon16, Eracleous2012, decarli13, Ju2013, liu13, shen13, Runnoe2015, Runnoe2017, li16, wang17}. Two drawbacks of this approach however remain: (a) temporal baselines of spectroscopic campaigns cover only a fraction of the SMBH binary orbital cycle and (b) intrinsic variability of the line profiles may mimic or hide radial velocity variations, making these measurements more challenging and less deterministic \citep{Runnoe2017}.

\section{Example ngVLA Studies}\label{sec:ngvla}
Dual and binary SMBH studies can be realized by a number of specific studies enabled by the ngVLA:
\begin{description}
\item[Blind surveys.] The ngVLA can survey many sources (thousands) with high sensitivity, high dynamic range, and high two-dimensional image fidelity (i.\,e.\ relatively complete $u, v$ coverage) in the frequency range 1--50\,GHz with moderate resolution. These abilities would enable a blind survey for resolved dual cores in the $\gtrsim$few parsec separation regime. Such a survey could also reveal single, offset cores that may arise from a gravitational-wave recoil kick following a SMBH merger. Candidates can be cross-matched with future optical/near-IR or X-ray (\eg, Athena) surveys to measure spectra, redshifts, and disturbed host galaxy structures.  While the multi-wavelength observations might not have the resolution of the \hbox{ngVLA}, they would provide a more complete understanding of a candidate and whether it
is a genuine dual \hbox{AGN}, a gravitational lens, 
a chance projection, or even a recoiling SMBH.
\item[Targeted surveys to test candidate binaries.] 
Many binary SMBH candidates are likely to be identified in the coming decades by numerous instruments (including PTAs, LSST and the extremely large optical telescopes currently being planned). The ngVLA will be the premier instrument to follow up these candidates in radio, with the goal of understanding their long-term dynamics through large-scale jet morphologies (at low GHz frequencies, given sufficient sensitivity to low surface brightness emission). A VLBI component on the ngVLA will enable observation and tracking of resolved dual cores, which will truly confirm a binary and allow in-depth study of SMBH binary evolution.
\item[Core variability.] The frequency range of ngVLA makes it well-suited to observe AGN that are dominated by core emission.
Past work has shown that in blind surveys at $\sim$10\,GHz, radio cores are more dominant than diffuse lobes \citep{massardi+11}. Thus, the high sensitivity of the ngVLA at tens of GHz frequencies can permit easier time-resolved radio quasar light curve monitoring, and thus have the potential to identify more examples that might exhibit periodic activity due to variable accretion or orbital precession.
\item[Multi-messenger searches.] 
GW experiments can only poorly localize their detections \citep{ellis13}. In concert with multi-wavelength facilities, the ngVLA could perform efficient follow-up surveys for all galaxies within a given error radius, identifying a host by looking for the signatures described above. With sufficiently high sensitivity on long baselines, the ngVLA could observe and track the orbital motions of radio cores for the most nearby GW candidates, thus enabling specific multi-messenger studies. In particular, PTAs will preferentially detect the most massive ($M_{\rm BH} \ga 10^9$ M$_{\odot}$), nearby ($z \la 0.1$) objects that are accessible to Earth-based VLBI \citep[\eg][]{kelley+17,mingarelli-2mass}.

\item[Long-baseline benefits.] 
A VLBI component to the ngVLA has clear benefits for this field of research. A long-baseline extension to the ngVLA, where each long baseline consists of a cluster of $\sim$5 antennas, would yield several major benefits. First, much of the imaging of double black holes is resolution-limited, meaning that going to longer baselines would immediately allow the probing of closer pairs of black holes. Second, long baselines would allow the maser experiments described in Section\,\ref{sec:masers}. Third, the clusters themselves can be used for pulsar timing when the ngVLA core is not being used as part of a VLBI project, giving access to the pulsars too far north for SKA timing.

\end{description}

\noindent The above studies enabled by the ngVLA will provide unique probes of SMBH binary evolution in regimes inaccessible with other methods. This includes crucial data on binaries in the truly gravitationally bound phase, such as information about their separations, inspiral timescales, luminosities and variability even when not in a rapidly-accreting ``quasar" state. Moreover, studies of radio jet morphology and energetics will provide unprecedented constraints on the mechanisms by which SMBH feedback drives host evolution and contributes to the origin of the observed SMBH-galaxy correlations. In concert with complementary SMBH binary studies by PTAs and LISA,\footnote{See also the chapter on ngVLA/LISA synergies in this volume; Ravi et al.} the ngVLA will be a key cornerstone in the coming era of multimessenger astronomy. The synergy between these electromagnetic and gravitational-wave probes of the binary regime has the potential to unravel the full puzzle of binary SMBH evolution and its close ties to the evolution of galaxies.


\acknowledgements 
SBS is supported by NSF award \#1458952. SBS and TJWL are members of the NANOGrav Physics Frontiers Center, which is supported by NSF award \#1430284. T.B. acknowledges support by the National Aeronautics and Space Administration under Grant No. NNX15AK84G issued through the Astrophysics Theory Program and by the Research Corporation for Science Advancement through a Cottrell Scholar Award. L.B. is supported by NSF award \#1715413. D. P. acknowledges support from the NSF through the Grote Reber Fellowship Program administered by Associated Universities, Inc./National Radio Astronomy Observatory. Y.S. acknowledges support from an Alfred P. Sloan Research Fellowship and NSF grant AST-1715579. Part of this research was carried out at the Jet Propulsion Laboratory, California Institute of Technology, under a contract with the National Aeronautics and Space Administration.


\bibliography{thebibliography}

\begin{thebibliography}{}
\expandafter\ifx\csname natexlab\endcsname\relax\def\natexlab#1{#1}\fi
\expandafter\ifx\csname url\endcsname\relax
  \def\url#1{\texttt{#1}}\fi
\expandafter\ifx\csname urlprefix\endcsname\relax\def\urlprefix{URL }\fi
\providecommand{\eprint}[2][]{\url{#2}}

\bibitem[{{Amaro-Seoane} et~al.(2017){Amaro-Seoane}, {Audley}, {Babak}, \&
  et~al.}]{as+17}
{Amaro-Seoane}, P., {Audley}, H., {Babak}, S., \& et~al. 2017, ArXiv e-prints.
  \eprint{1702.00786}

\bibitem[{{Bansal} et~al.(2017){Bansal}, {Taylor}, {Peck}, {Zavala}, \&
  {Romani}}]{bansal}
{Bansal}, K., {Taylor}, G.~B., {Peck}, A.~B., {Zavala}, R.~T., \& {Romani},
  R.~W. 2017, \apj, 843, 14

\bibitem[{{Barrows} et~al.(2016){Barrows}, {Comerford}, {Greene}, \&
  {Pooley}}]{barrows16}
{Barrows}, R.~S., {Comerford}, J.~M., {Greene}, J.~E., \& {Pooley}, D. 2016,
  \apj, 829, 37. \eprint{1606.01253}

\bibitem[{{Barth} et~al.(2015){Barth}, {Bennert}, {Canalizo}, \&
  et~al.}]{barth15}
{Barth}, A.~J., {Bennert}, V.~N., {Canalizo}, G., \& et~al. 2015, \apjs, 217,
  26. \eprint{1503.01146}

\bibitem[{{Begelman} et~al.(1980){Begelman}, {Blandford}, \&
  {Rees}}]{begelman80}
{Begelman}, M.~C., {Blandford}, R.~D., \& {Rees}, M.~J. 1980, \nat, 287, 307

\bibitem[{{Blandford} \& {K{\"o}nigl}(1979)}]{blandford+79}
{Blandford}, R.~D., \& {K{\"o}nigl}, A. 1979, \apj, 232, 34

\bibitem[{{Blecha} et~al.(2013){Blecha}, {Loeb}, \& {Narayan}}]{blecha13}
{Blecha}, L., {Loeb}, A., \& {Narayan}, R. 2013, \mnras, 429, 2594.
  \eprint{1201.1904}

\bibitem[{{Bogdanovi{\'c}}(2015)}]{Bogdanovic2015}
{Bogdanovi{\'c}}, T. 2015, in Gravitational Wave Astrophysics, edited by C.~F.
  {Sopuerta}, vol.~40 of Astrophysics and Space Science Proceedings, 103.
  \eprint{1406.5193}

\bibitem[{{Bogdanovi{\'c}} et~al.(2009){Bogdanovi{\'c}}, {Eracleous}, \&
  {Sigurdsson}}]{bogdanovic09}
{Bogdanovi{\'c}}, T., {Eracleous}, M., \& {Sigurdsson}, S. 2009, New Astronomy
  reviews, 53, 113

\bibitem[{{Bon} et~al.(2012){Bon}, {Jovanovi{\'c}}, {Marziani}, \&
  et~al.}]{bon12}
{Bon}, E., {Jovanovi{\'c}}, P., {Marziani}, P., \& et~al. 2012, \apj, 759, 118.
  \eprint{1209.4524}

\bibitem[{{Bon} et~al.(2016){Bon}, {Zucker}, {Netzer}, \& et~al.}]{bon16}
{Bon}, E., {Zucker}, S., {Netzer}, H., \& et~al. 2016, \apjs, 225, 29.
  \eprint{1606.04606}

\bibitem[{{Braatz} et~al.(2010){Braatz}, {Reid}, {Humphreys}, {Henkel},
  {Condon}, \& {Lo}}]{Braatz10}
{Braatz}, J.~A., {Reid}, M.~J., {Humphreys}, E.~M.~L., {Henkel}, C., {Condon},
  J.~J., \& {Lo}, K.~Y. 2010, \apj, 718, 657. \eprint{1005.1955}

\bibitem[{{Burke-Spolaor}(2011)}]{radiocensus}
{Burke-Spolaor}, S. 2011, \mnras, 410, 2113. \eprint{1008.4382}

\bibitem[{{Burke-Spolaor}(2013)}]{sbs-cqg}
--- 2013, Classical and Quantum Gravity, 30, 224013. \eprint{1308.4408}

\bibitem[{{Burke-Spolaor} et~al.(2018){Burke-Spolaor}, Hazboun, Kelley, Lazio,
  Madison, McMann, Mingarelli, Rasskazov, Siemens, Simon, Smith, \&
  Taylor}]{GWAWG-paper}
{Burke-Spolaor}, S., Hazboun, J., Kelley, L., Lazio, J., Madison, D., McMann,
  N., Mingarelli, C., Rasskazov, A., Siemens, X., Simon, J., Smith, T., \&
  Taylor, S. 2018, in prep.

\bibitem[{{Charisi} et~al.(2016){Charisi}, {Bartos}, {Haiman}, \&
  et~al.}]{charisi16}
{Charisi}, M., {Bartos}, I., {Haiman}, Z., \& et~al. 2016, \mnras, 463, 2145.
  \eprint{1604.01020}

\bibitem[{{Charisi} et~al.(2018){Charisi}, {Haiman}, {Schiminovich}, \&
  {D'Orazio}}]{charisi+18}
{Charisi}, M., {Haiman}, Z., {Schiminovich}, D., \& {D'Orazio}, D.~J. 2018,
  \mnras. \eprint{1801.06189}

\bibitem[{{Comerford} et~al.(2015){Comerford}, {Pooley}, {Barrows}, \&
  et~al.}]{comerford15}
{Comerford}, J.~M., {Pooley}, D., {Barrows}, R.~S., \& et~al. 2015, \apj, 806,
  219. \eprint{1504.01391}

\bibitem[{{Condon} et~al.(2011){Condon}, {Darling}, {Kovalev}, \&
  {Petrov}}]{condon-2mass}
{Condon}, J., {Darling}, J., {Kovalev}, Y.~Y., \& {Petrov}, L. 2011, ArXiv
  e-prints. \eprint{1110.6252}

\bibitem[{{Deane} et~al.(2014){Deane}, {Paragi}, {Jarvis}, \& et~al.}]{deane}
{Deane}, R.~P., {Paragi}, Z., {Jarvis}, M.~J., \& et~al. 2014, \nat, 511, 57

\bibitem[{{Decarli} et~al.(2013){Decarli}, {Dotti}, {Fumagalli}, \&
  et~al.}]{decarli13}
{Decarli}, R., {Dotti}, M., {Fumagalli}, M., \& et~al. 2013, \mnras, 433, 1492.
  \eprint{1305.4941}

\bibitem[{{Deller} \& {Middelberg}(2014)}]{deller+middelberg}
{Deller}, A.~T., \& {Middelberg}, E. 2014, \aj, 147, 14. \eprint{1310.8191}

\bibitem[{{D'Orazio} \& {Loeb}(2017)}]{dorazioloeb}
{D'Orazio}, D.~J., \& {Loeb}, A. 2017, ArXiv e-prints. \eprint{1712.02362}

\bibitem[{{Ellis}(2013)}]{ellis13}
{Ellis}, J.~A. 2013, Classical and Quantum Gravity, 30, 224004.
  \eprint{1305.0835}

\bibitem[{{Eracleous} et~al.(2012){Eracleous}, {Boroson}, {Halpern}, \&
  {Liu}}]{Eracleous2012}
{Eracleous}, M., {Boroson}, T.~A., {Halpern}, J.~P., \& {Liu}, J. 2012, \apjs,
  201, 23. \eprint{1106.2952}

\bibitem[{{Favata}(2009)}]{favata09}
{Favata}, M. 2009, \apjl, 696, L159. \eprint{0902.3660}

\bibitem[{{Fu} et~al.(2011){Fu}, {Zhang}, {Assef}, \& et~al.}]{fu11}
{Fu}, H., {Zhang}, Z.-Y., {Assef}, R.~J., \& et~al. 2011, \apjl, 740, L44+.
  \eprint{1109.0008}

\bibitem[{{Gao} et~al.(2016){Gao}, {Braatz}, {Reid}, {Lo}, {Condon}, {Henkel},
  {Kuo}, {Impellizzeri}, {Pesce}, \& {Zhao}}]{Gao16}
{Gao}, F., {Braatz}, J.~A., {Reid}, M.~J., {Lo}, K.~Y., {Condon}, J.~J.,
  {Henkel}, C., {Kuo}, C.~Y., {Impellizzeri}, C.~M.~V., {Pesce}, D.~W., \&
  {Zhao}, W. 2016, \apj, 817, 128. \eprint{1511.08311}

\bibitem[{{Gaskell}(1983)}]{gaskell83}
{Gaskell}, C.~M. 1983, in Liege International Astrophysical Colloquia, edited
  by {J.-P.~Swings}, vol.~24, 473

\bibitem[{{Gaskell}(1996)}]{gaskell96}
--- 1996, \apjl, 464, L107. \eprint{astro-ph/9605185}

\bibitem[{{Godfrey} et~al.(2012){Godfrey}, {Lovell}, {Burke-Spolaor}, \&
  et~al.}]{godfrey_0637-752}
{Godfrey}, L.~E.~H., {Lovell}, J.~E.~J., {Burke-Spolaor}, S., \& et~al. 2012,
  \apjl, 758, L27. \eprint{1209.4637}

\bibitem[{{Graham} et~al.(2015){Graham}, {Djorgovski}, {Stern}, \&
  et~al.}]{Graham2015}
{Graham}, M.~J., {Djorgovski}, S.~G., {Stern}, D., \& et~al. 2015, \mnras, 453,
  1562. \eprint{1507.07603}

\bibitem[{{Helmboldt} et~al.(2007){Helmboldt}, {Taylor}, {Tremblay}, \&
  et~al.}]{VIPS}
{Helmboldt}, J.~F., {Taylor}, G.~B., {Tremblay}, S., \& et~al. 2007, \apj, 658,
  203. \eprint{astro-ph/0611459}

\bibitem[{{Hopkins}(2012)}]{hopkins12}
{Hopkins}, P.~F. 2012, \mnras, 420, L8. \eprint{1101.4230}

\bibitem[{{Hopkins} et~al.(2014){Hopkins}, {Kocevski}, \& {Bundy}}]{hopkins14}
{Hopkins}, P.~F., {Kocevski}, D.~D., \& {Bundy}, K. 2014, \mnras, 445, 823.
  \eprint{1309.6321}

\bibitem[{{Ju} et~al.(2013){Ju}, {Greene}, {Rafikov}, \& et~al.}]{Ju2013}
{Ju}, W., {Greene}, J.~E., {Rafikov}, R.~R., \& et~al. 2013, \apj, 777, 44.
  \eprint{1306.4987}

\bibitem[{{Kelley} et~al.(2017{\natexlab{a}}){Kelley}, {Blecha}, \&
  {Hernquist}}]{kelley+17}
{Kelley}, L.~Z., {Blecha}, L., \& {Hernquist}, L. 2017{\natexlab{a}}, \mnras,
  464, 3131. \eprint{1606.01900}

\bibitem[{{Kelley} et~al.(2017{\natexlab{b}}){Kelley}, {Blecha}, {Hernquist},
  {Sesana}, \& {Taylor}}]{kelley+submitted}
{Kelley}, L.~Z., {Blecha}, L., {Hernquist}, L., {Sesana}, A., \& {Taylor}, S.
  2017{\natexlab{b}}, submitted. \eprint{1711.00075}

\bibitem[{{Kharb} et~al.(2017){Kharb}, {Lal}, \& {Merritt}}]{kharb+17}
{Kharb}, P., {Lal}, D.~V., \& {Merritt}, D. 2017, Nature Astronomy, 1, 727.
  \eprint{1709.06258}

\bibitem[{{Klein} et~al.(2016){Klein}, {Barausse}, {Sesana}, {Petiteau},
  {Berti}, {Babak}, {Gair}, {Aoudia}, {Hinder}, {Ohme}, \&
  {Wardell}}]{lisa-science-2}
{Klein}, A., {Barausse}, E., {Sesana}, A., {Petiteau}, A., {Berti}, E.,
  {Babak}, S., {Gair}, J., {Aoudia}, S., {Hinder}, I., {Ohme}, F., \&
  {Wardell}, B. 2016, \prd, 93, 024003. \eprint{1511.05581}

\bibitem[{{Komossa} et~al.(2003){Komossa}, {Burwitz}, {Hasinger}, \&
  et~al.}]{ngc6240}
{Komossa}, S., {Burwitz}, V., {Hasinger}, G., \& et~al. 2003, \apjl, 582, L15.
  \eprint{arXiv:0212099}

\bibitem[{{Koss} et~al.(2011){Koss}, {Mushotzky}, {Treister}, \&
  et~al.}]{koss11}
{Koss}, M., {Mushotzky}, R., {Treister}, E., \& et~al. 2011, \apjl, 735, L42.
  \eprint{1106.2163}

\bibitem[{{Koss} et~al.(2016){Koss}, {Glidden}, {Balokovi{\'c}}, \&
  et~al.}]{koss16}
{Koss}, M.~J., {Glidden}, A., {Balokovi{\'c}}, M., \& et~al. 2016, \apjl, 824,
  L4. \eprint{1708.06762}

\bibitem[{{Kuo} et~al.(2011){Kuo}, {Braatz}, {Condon}, {Impellizzeri}, {Lo},
  {Zaw}, {Schenker}, {Henkel}, {Reid}, \& {Greene}}]{Kuo11}
{Kuo}, C.~Y., {Braatz}, J.~A., {Condon}, J.~J., {Impellizzeri}, C.~M.~V., {Lo},
  K.~Y., {Zaw}, I., {Schenker}, M., {Henkel}, C., {Reid}, M.~J., \& {Greene},
  J.~E. 2011, \apj, 727, 20. \eprint{1008.2146}

\bibitem[{{Kuo} et~al.(2013){Kuo}, {Braatz}, {Reid}, {Lo}, {Condon},
  {Impellizzeri}, \& {Henkel}}]{Kuo13}
{Kuo}, C.~Y., {Braatz}, J.~A., {Reid}, M.~J., {Lo}, K.~Y., {Condon}, J.~J.,
  {Impellizzeri}, C.~M.~V., \& {Henkel}, C. 2013, \apj, 767, 155.
  \eprint{1207.7273}

\bibitem[{{Li} et~al.(2016){Li}, {Wang}, {Ho}, \& et~al.}]{li16}
{Li}, Y.-R., {Wang}, J.-M., {Ho}, L.~C., \& et~al. 2016, \apj, 822, 4.
  \eprint{1602.05005}

\bibitem[{{Liu} et~al.(2016){Liu}, {Gezari}, {Burgett}, \& et~al.}]{liu16}
{Liu}, T., {Gezari}, S., {Burgett}, W., \& et~al. 2016, \apj, 833, 6.
  \eprint{1609.09503}

\bibitem[{{Liu} et~al.(2018){Liu}, {Gezari}, \& {Miller}}]{liu18}
{Liu}, T., {Gezari}, S., \& {Miller}, M.~C. 2018, ArXiv e-prints.
  \eprint{1803.05448}

\bibitem[{{Liu} et~al.(2013){Liu}, {Civano}, {Shen}, {Green}, {Greene}, \&
  {Strauss}}]{liu13b}
{Liu}, X., {Civano}, F., {Shen}, Y., {Green}, P., {Greene}, J.~E., \&
  {Strauss}, M.~A. 2013, \apj, 762, 110

\bibitem[{{Liu} et~al.(2014){Liu}, {Shen}, {Bian}, {Loeb}, \&
  {Tremaine}}]{liu13}
{Liu}, X., {Shen}, Y., {Bian}, F., {Loeb}, A., \& {Tremaine}, S. 2014, \apj,
  789, 140. \eprint{1312.6694}

\bibitem[{{Liu} et~al.(2011){Liu}, {Shen}, {Strauss}, \& {Hao}}]{liu+11}
{Liu}, X., {Shen}, Y., {Strauss}, M.~A., \& {Hao}, L. 2011, \apj, 737, 101.
  \eprint{1104.0950}

\bibitem[{{Massardi} et~al.(2011){Massardi}, {Ekers}, {Murphy}, \&
  et~al.}]{massardi+11}
{Massardi}, M., {Ekers}, R.~D., {Murphy}, T., \& et~al. 2011, \mnras, 412, 318

\bibitem[{{Merritt} \& {Ekers}(2002)}]{merrittekers}
{Merritt}, D., \& {Ekers}, R.~D. 2002, Science, 297, 1310.
  \eprint{astro-ph/0208001}

\bibitem[{{Merritt} \& {Milosavljevi{\'c}}(2005)}]{merrittLRR}
{Merritt}, D., \& {Milosavljevi{\'c}}, M. 2005, Living Reviews in Relativity,
  8. \eprint{astro-ph/0410364}

\bibitem[{{Mingarelli} et~al.(2017){Mingarelli}, {Lazio}, {Sesana}, \&
  et~al.}]{mingarelli-2mass}
{Mingarelli}, C.~M.~F., {Lazio}, T.~J.~W., {Sesana}, A., \& et~al. 2017, Nature
  Astronomy, 1, 886

\bibitem[{{Miyoshi} et~al.(1995){Miyoshi}, {Moran}, {Herrnstein}, {Greenhill},
  {Nakai}, {Diamond}, \& {Inoue}}]{Miyoshi95}
{Miyoshi}, M., {Moran}, J., {Herrnstein}, J., {Greenhill}, L., {Nakai}, N.,
  {Diamond}, P., \& {Inoue}, M. 1995, \nat, 373, 127

\bibitem[{{M{\"u}ller-S{\'a}nchez} et~al.(2015){M{\"u}ller-S{\'a}nchez},
  {Comerford}, {Nevin}, \& et~al.}]{mullersanchez15}
{M{\"u}ller-S{\'a}nchez}, F., {Comerford}, J.~M., {Nevin}, R., \& et~al. 2015,
  \apj, 813, 103. \eprint{1509.04291}

\bibitem[{{Popovi{\'c}}(2012)}]{popovic12}
{Popovi{\'c}}, L.~{\v C}. 2012, New Astronomy Reviews, 56, 74.
  \eprint{1109.0710}

\bibitem[{{Roberts} et~al.(2018){Roberts}, {Saripalli}, {Wang}, \&
  et~al.}]{roberts18}
{Roberts}, D.~H., {Saripalli}, L., {Wang}, K.~X., \& et~al. 2018, \apj, 852,
  47. \eprint{1708.02306}

\bibitem[{{Rodriguez} et~al.(2006){Rodriguez}, {Taylor}, {Zavala}, {Peck},
  {Pollack}, \& {Romani}}]{rodriguez}
{Rodriguez}, C., {Taylor}, G., {Zavala}, R., {Peck}, A., {Pollack}, L., \&
  {Romani}, R. 2006, \apj, 646, 49

\bibitem[{{Roos} et~al.(1993){Roos}, {Kaastra}, \& {Hummel}}]{roos+93}
{Roos}, N., {Kaastra}, J.~S., \& {Hummel}, C.~A. 1993, \apj, 409, 130

\bibitem[{{Runnoe} et~al.(2015){Runnoe}, {Eracleous}, {Mathes}, \&
  et~al.}]{Runnoe2015}
{Runnoe}, J.~C., {Eracleous}, M., {Mathes}, G., \& et~al. 2015, \apjs, 221, 7.
  \eprint{1509.02575}

\bibitem[{{Runnoe} et~al.(2017){Runnoe}, {Eracleous}, {Pennell}, \&
  et~al.}]{Runnoe2017}
{Runnoe}, J.~C., {Eracleous}, M., {Pennell}, A., \& et~al. 2017, \mnras, 468,
  1683. \eprint{1702.05465}

\bibitem[{{Saripalli} \& {Subrahmanyan}(2009)}]{xnotbinary}
{Saripalli}, L., \& {Subrahmanyan}, R. 2009, \apj, 695, 156. \eprint{0811.1907}

\bibitem[{{Sesana} et~al.(2017){Sesana}, {Haiman}, {Kocsis}, \&
  {Kelley}}]{sesana17}
{Sesana}, A., {Haiman}, Z., {Kocsis}, B., \& {Kelley}, L.~Z. 2017, ArXiv
  e-prints. \eprint{1703.10611}

\bibitem[{{Shen} et~al.(2013){Shen}, {Liu}, {Loeb}, \& {Tremaine}}]{shen13}
{Shen}, Y., {Liu}, X., {Loeb}, A., \& {Tremaine}, S. 2013, \apj, 775, 49.
  \eprint{1306.4330}

\bibitem[{{Tremblay} et~al.(2016){Tremblay}, {Taylor}, {Ortiz}, \&
  et~al.}]{tremblay16}
{Tremblay}, S.~E., {Taylor}, G.~B., {Ortiz}, A.~A., \& et~al. 2016, \mnras,
  459, 820. \eprint{1603.03094}

\bibitem[{{Valtonen} et~al.(2008){Valtonen}, {Lehto}, {Nilsson}, \&
  et~al.}]{valtonen08}
{Valtonen}, M.~J., {Lehto}, H.~J., {Nilsson}, K., \& et~al. 2008, \nat, 452,
  851. \eprint{0809.1280}

\bibitem[{{Vaughan} et~al.(2016){Vaughan}, {Uttley}, {Markowitz}, \&
  et~al.}]{vaughan+16}
{Vaughan}, S., {Uttley}, P., {Markowitz}, A.~G., \& et~al. 2016, \mnras, 461,
  3145. \eprint{1606.02620}

\bibitem[{{Wang} et~al.(2017){Wang}, {Greene}, {Ju}, {Rafikov}, {Ruan}, \&
  {Schneider}}]{wang17}
{Wang}, L., {Greene}, J.~E., {Ju}, W., {Rafikov}, R.~R., {Ruan}, J.~J., \&
  {Schneider}, D.~P. 2017, \apj, 834, 129. \eprint{1611.00039}

\end{thebibliography}

\end{document}